# How to Report and Benchmark Emerging Field-Effect Transistors

Zhihui Cheng[1,2,†], Chin-Sheng Pang[2], Peiqi Wang[3], Son T. Le[1,4], Yanqing Wu[5], Davood Shahrjerdi[6,7], Iuliana Radu[8], Max C. Lemme[9,10], Lian-Mao Peng[11], Xiangfeng Duan[3], Zhihong Chen[2], Joerg Appenzeller[2], Steven J. Koester[12], Eric Pop[13], Aaron D. Franklin[14,15,†], Curt A. Richter[1,†]

**The use of organic, oxide and low-dimensional materials in field-effect transistors has now been studied for decades. However, properly reporting and comparing device performance remains challenging due to the interdependency of multiple device parameters. The interdisciplinarity of this research community has also led to a lack of consistent reporting and benchmarking guidelines. Here we propose guidelines for reporting and benchmarking key field-effect transistor parameters and performance metrics. We provide an example of this reporting and benchmarking process by using a two-dimensional semiconductor field-effect transistor. Our guidelines should help promote an improved approach for assessing device performance in emerging field-effect transistors, helping the field to progress in a more consistent and meaningful way.**

Research into field-effect transistors (FETs) based on emerging nanomaterials, including carbon nanotubes[1,2], graphene[3], phosphorene[4], silicene[5], tellurene[6], transition metal dichalcogenides[7–9], organic semiconductors[10,11] and ultrathin metal oxides[12], is thriving. Such studies allow the fundamental properties of the materials to be explored, and may lead to the development of various commercial applications; however, effectively and uniformly assessing the performance of emergent FETs is difficult due to the dependence of performance metrics on unique aspects of the device structure (Fig. 1a)[13].

Structural parameters that influence device performance include channel ($L_\text{ch}$) and contact lengths ($L_\text{c}$), gate insulator thickness ($t_\text{ins}$) and permittivity ($\epsilon_\text{ins}$), contact metal types, the thickness of channel material ($t_\text{ch}$), and gating scheme (e.g., top, bottom, gate-all-around, multi-channel). Performance metrics include on-current ($I_\text{on}$), off-current ($I_\text{off}$), $I_\text{on}/I_\text{off}$ ratio, contact resistance ($R_\text{c}$), transconductance ($g_\text{m}$), subthreshold swing ($SS$), channel mobilities, and drain-induced barrier lowering (DIBL). While different studies reported in the literature often include some of these benchmarking figures, they struggle to capture the myriad variables, making comparisons inaccurate or even biased at times. In addition, the emerging device community consists of researchers from disparate disciplines — including electrical engineering, chemistry, materials science, and physics — which also makes consistent reporting and benchmarking challenging. In this Perspective, we examine the challenges involved in assessing the operation and performance of FETs based on emerging materials, and provide guidelines on how to report and benchmark the devices.

A full list of affiliations appears at the end of the paper.

**Field-effect transistor structure and key parameters**

In a FET, the structural parameters determine the electric fields and the eventual device performance (Fig. 1a). Subthreshold, transfer, and output characteristics in Figs. 1b-d are the most common *I-V* (current-voltage) curves to capture the device performance. Plotting the log of the drain current ($I_D$) as a function of gate-source voltage ($V_{GS}$) highlights the subthreshold (i.e., off-state) device behaviour. In contrast, transfer characteristics plot $I_D$ vs. $V_{GS}$ on a linear scale and emphasize the device behaviour after $V_{GS}$ exceeds the threshold voltage ($V_T$), where the device is in the on-state. Ideally, the gate leakage current ($I_G$) vs. $V_{GS}$ should be plotted on the subthreshold plot as well.

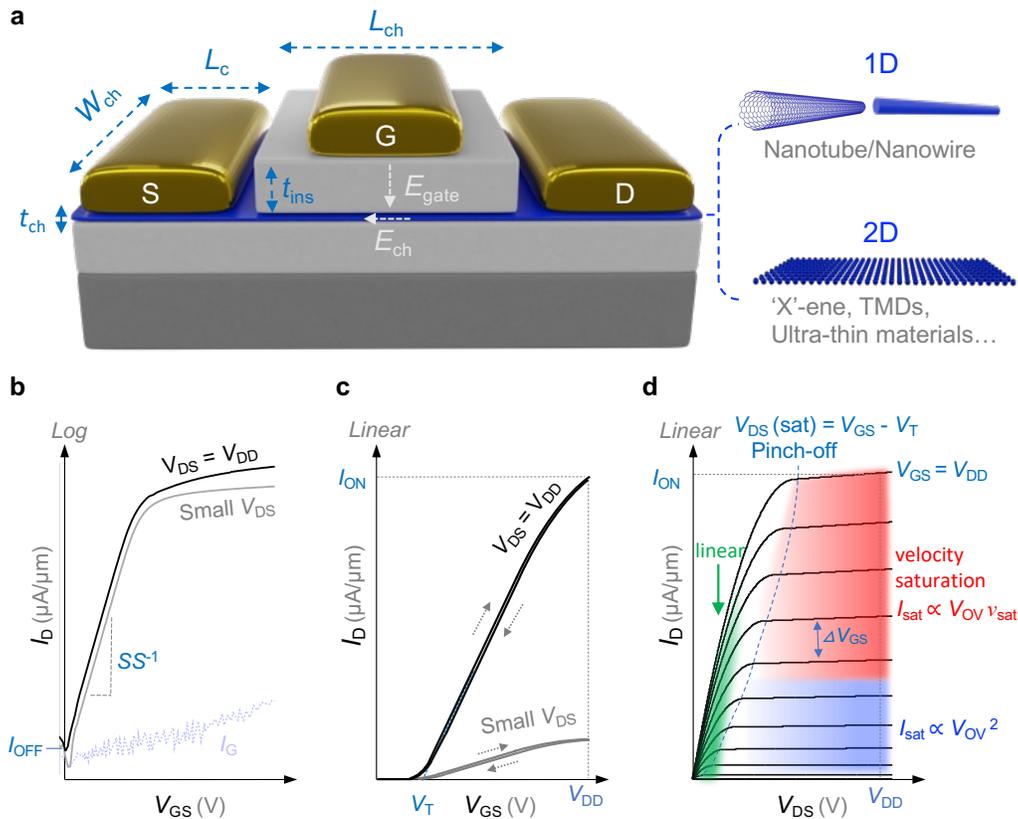

**Fig. 1 | Basic device structure and electrical characteristics**. **a**, Diagram of a typical nanomaterial-based n-type FET highlighting the structure parameters and electric fields. **b**, Subthreshold and **c**, transfer ($I_D$-$V_{GS}$) curves of an *n*-type FET under different $V_{DS}$ voltages. Representative curves are shown for single-sweep in (**b**), but forward- and backward-sweeps of $V_{GS}$ should be collected to determine hysteresis, as shown in (**c**). The gate leakage current is also shown in (**b**). **d**, Output ($I_D$-$V_{DS}$) curves of the device with three main operation regimes labelled.[14] $V_{GS}$ is swept from low to high in steps of $\Delta V_{GS}$. $v_{sat}$ is the saturation velocity of carriers in the channel material. $I_{sat}$ scales as $V_{OV}^2$ in the classical pinch-off regime, but only linearly with $V_{OV}$ when the velocity saturates. Self-heating could render this increase even sub-linear.[14,15] Note, $V_{DD}$ is the supply voltage for the transistor (i.e., the target maximum voltage of operation for both $V_{DS}$ and $V_{GS}$).

The $I_D$-$V_{GS}$ sweeps in Fig. 1b-c should be conducted at both "small" and "high" drain-source voltage ($V_{DS}$) values to characterize the device operation in both linear and saturation regimes. We note that the "small" $V_{DS}$ value should be sufficiently small to ensure linear regime operation, but greater than $\sim 2k_B T$ (where $k_B$ is the Boltzmann constant and $T$ is the absolute temperature, i.e., ~50 mV at room temperature) to ensure that the subthreshold behaviour, and here in particular DIBL ($\partial V_T/\partial V_{DS}$), is not misinterpreted due to thermal injection of carriers from the drain. These curves enable easy extraction of DIBL to demonstrate how $V_{DS}$ impacts $V_T$. The transfer characteristics should be acquired with forward and backward sweeps, checking for the presence of any hysteresis due to charge trapping[16]. When comparing hysteresis from different devices, the precise measurement conditions such as sweep rates, hold times, and maximum bias voltages should be listed as these parameters influence hysteresis. If hysteresis exists, it should be accounted for in the analysis of $V_T$ uncertainty and other device parameters that depend on $V_T$.

In the output characteristics (Fig. 1d), three main operation regimes are highlighted. The linear regime is characterized by the linear increase of $I_D$ with both $V_{DS}$ and $V_{GS}$. After $V_{DS}$ surpasses the overdrive voltage ($V_{OV} = V_{GS} - V_T$, for $n$-channel FETs), $I_D$ starts to saturate to $I_{sat}$, which could (based on the classical FET model[17]) increase quadratically with $V_{OV}$ in the pinch-off regime and linearly with $V_{OV}$ in the velocity saturation regime[14]. Note, the linear regime may present as nonlinear (often exponential) in the event of poor carrier injection at the contacts, such as from large Schottky barriers.

Multiple performance parameters can be extracted from the $I$-$V$ curves in Fig. 1b-d. The most important performance metrics are the currents, which must be reported normalized by the channel width, $W_{ch}$ (e.g., units of µA/µm). For 1D or quasi-1D devices, it is common to first report the current per CNT/nanowire/nanosheet stack. Then the current can be normalized to µA/µm by considering the expected channel density and pitch of the channel material (e.g., 10 µA/CNT with 50 CNTs/µm, giving 500 µA/µm) since the aerial footprint of the device is a critical aspect of performance. When extracting $I_{on}$ and $I_{off}$ from these $I$-$V$ curves, in a simplified scenario, $I_{off}$ is the $I_D$ measured at $V_{GS} = 0$ and $V_{DS} = V_{DD}$, whereas $I_{on}$ is the $I_D$ measured at $V_{GS} = V_{DS} = V_{DD}$. Here, $V_{DD}$ is the voltage that would be supplied to operate the transistors. (For mainstream silicon technology, $V_{DD}$ has dropped to 1 V near 2010 and to 0.7 V in recent years[18].) For modern technologies, the exact value of $V_{DD}$ depends on the application. For example, if the emergent transistor is used as an access transistor in a dynamic random-access memory (DRAM), then its $V_{DD}$ will be a small value to ensure linear regime operation in the on-state. Reported emergent devices often do not have threshold voltages tuned such that $V_{GS} = 0$ is a sensible off-state; additionally, there is often not a well-defined $V_{DD}$ value due to the wide variety of device structural parameters. We hence suggest extracting the maximum and minimum $I_D$ ($I_{max}$ and $I_{min}$) from a typical subthreshold curve and reporting the $I_{max}/I_{min}$ ratios when $V_{DS}$ is biased in both linear and saturation regimes. A more detailed description on reporting and benchmarking $I_{max}/I_{min}$ is in Note S1.

When reporting the $I_{sat}$ of a device, it is necessary to note the carrier density $n$, at which the $I_{sat}$ is extracted. Ideally, the Hall effect is used to measure the carrier density for the channel material, but for most researchers in the FET community, more accessible approaches are needed that do not require specially designed test structures. In the linear regime, the average carrier density can be estimated as $n \approx C_{ins}(V_{OV} - V_{DS}/2)/q$, where $C_{ins}$ is the gate insulator capacitance and $q$ is the elementary charge; however, in the saturation regime, the depletion region in the channel complicates the estimation. The carrier density *near the source side* is the same for both the linear and saturation regimes. For convenience and simplicity, we recommend clearly labelling the carrier density near the source as $n_S$ ($=C_{ins}V_{OV}/q$) and using this value for both operation regimes. To determine $V_{OV}$, $V_T$ is usually estimated using extrapolation in the linear portion of the transfer curve, as listed in Table 1. Other methods, such as constant current[19], Y-function methods[20,21] and four-probe measurements[22], can be used to cross-check the linear extraction of $V_T$ and reduce the variation when estimating $n_S$. More discussion regarding $V_T$ extraction is given in Note S2.

In addition to $I_D$, $R_c$ is also essential to represent device performance. The transfer length method (TLM) is the most commonly used approach for extracting $R_c$, along with the sheet resistance, $R_{sh}$, of the channel (in units of Ω/square)[23]. The TLM approach requires a series of FETs with different channel lengths and consistent contact and gating configurations. It entails plotting the total resistance of each device versus $L_{ch}$ at a given $n_S$, allowing $R_c$ to be extracted as the extrapolated y-axis intercept from a linear fit to the data points. Typically, the $V_{DS}$ for calculating the total resistance is the "small" $V_{DS}$ used in Fig. 1b to ensure linear regime operation. The channel lengths in the TLM should range from "short" (where the total resistance is dominated by $R_c$) to "long" (dominated by channel resistance, $R_{ch} = R_{tot} - 2R_c$ or $R_{sh}L_{ch}$) where the actual "short" and "long" channel lengths will depend on the relationship between the channel resistance and the contact resistance. A more detailed discussion on extracting $R_c$ and other considerations using TLM data is in Note S3.

Another frequently reported parameter is the carrier mobility of the channel material. Among various forms of mobility, the field-effect mobility $\mu_{FE} = L_{ch}g_m/(W_{ch}C_{ins}V_{DS})$ is often used. However, $\mu_{FE}$ can be underestimated[21,24] or overestimated[22,25,26] relative to the drift mobility of the channel material depending on the details of $V_{DS}$, $V_{GS}$, $R_c$, $L_{ch}$, and gate capacitance. In particular, gated contact effects can significantly affect mobility extraction. Although different approaches[21,23] have been proposed to make $\mu_{FE}$ less dependent on various factors, such as $R_c$ and $L_{ch}$, none of them is sufficiently general enough to be widely adopted. Conductivity mobility ($\mu_{con}$) has the advantage of strictly reflecting the channel material properties and the quality of the channel-dielectric interface[23,27]. In a FET, $\mu_{con}$[23] can be estimated from the sheet resistance of the semiconductor channel and the carrier density $n_S$ (see Table 1); thus, it does not involve the contact resistance or the device structure. High mobility is often a goal for research FETs; when such reports are made, it is critical to clearly state how the values are determined, and ideally multiple approaches (such as $\mu_{FE}$ and four-probe measurements[22]) are taken to cross-validate the claims. It is worth noting that the usefulness of channel mobility as an indicator of performance in aggressively scaled FETs is debatable as devices with channel lengths < 30 nm are going to be strongly limited by contact resistance (including carrier injection efficiency) with minimal dependence on transport in the channel[28,29].

The most representative FET parameters are listed in Table 1 as a suggested reporting checklist. Additional parameters are briefly discussed in Note S4.

**Table 1| Checklist of suggested device parameters to report**

| Name | Characteristics | Additional details |
|---|---|---|
| Structural parameters | Contact length, $L_c$<br>Channel length, $L_{ch}$<br>Channel width, $W_{ch}$<br>Insulator thickness, $t_{ins}$<br>Channel thickness, $t_{ch}$ | Specify contact and gating geometry/materials; include high-resolution electron microscopy evidence when reporting sub-20 nm dimensions (especially for $L_c$ and $L_{ch}$) |
| Insulator capacitance, $C_{ins}$ | Capacitance-voltage or Capacitance-frequency | Measured $C_{ins}$ is more accurate than estimating $\epsilon_{ins}$ especially when a high-$k$ insulator is used |
| Threshold voltage, $V_T$, and hysteresis, $\Delta V_T$ | Extrapolation in the linear portion of the transfer curve[19] | • $I_D$-$V_{GS}$ should have forward and backward sweeps<br>• Consider $V_T$ uncertainty due to hysteresis (charge trapping), the dependence of $V_{DS}$ and $I$-$V$ sweeps (Note S2) |
| Drain current in saturation regime, $I_{sat}$ | $I_D$-$V_{DS}$ (saturation regime) | • Sweep $I_D$-$V_{DS}$ to saturation regime<br>• Specify carrier density where $I_{sat}$ is extracted<br>• Normalized by channel width |
| Contact resistance, $R_c$ | TLM[23] (Note S3) | • Linear regime (small $V_{DS}$)<br>• Specify carrier density $n_S$ or plot $R_c$ vs. $n_S$<br>• TLM should have at least four channels and include at least one each of contact and channel resistance dominated devices |
| Conductivity mobility, $\mu_{con}$ | $\dfrac{1}{qn_S R_{sh}}$ | • $R_{sh}$ is extracted from the slope of TLM plots or from four-probe measurements[22]. Unit: Ω/square<br>• Carrier density near the source: $n_S \approx C_{ins}V_{OV}/q$<br>• Mobility from $R_{sh} = (qn_S\mu)^{-1}$<br>• Plot mobility vs $n_S$ to show field dependence |

| Transconductance, $g_m$ | Transfer or output curves $g_m = \frac{\partial I_D}{\partial V_{GS}}$ at certain $V_{DS}$ | Specify $g_m$ (linear) or $g_m$ (saturation) |
|---|---|---|
| Subthreshold Swing, SS | Subthreshold curves (inverse slope in mV/decade below $V_T$) | • SS depends on $C_{ins}$ and interface trap capacitance $C_{it}$<br>• Plot SS vs. $\log_{10}(I_D)$ |
| $I_{on}/I_{off}$ | Subthreshold curves at saturation regime, $V_{DS} = V_{DS}$ (sat) | • Report $I_{max}/I_{min}$ as an alternative along with $n_S$ range<br>• Plot $I_G$ vs $V_{GS}$ to show leakage current |
| DIBL | $\Delta V_T / \Delta V_{DS}$ from transfer curves | Key for short-channel devices |

Beyond the parameters in Table 1, showing statistics and variation is strongly encouraged to obtain comprehensive coverage of the device performance. The variation can be shown as error bars, box plots, coefficient of variation, or cumulative distribution function (see Note S5 for demonstration). Due to many nonidealities associated with emerging materials or unconventional device geometries, it is almost unavoidable that there could be considerable uncertainties in many extracted parameters, including $R_c$, $n_S$, and mobilities. These parameters are often interdependent. Reducing device variation is a major research theme for the eventual application of emergent FETs. Whatever measurements and specific analysis approaches are taken to determine these parameters, the details should be clearly and explicitly reported, and our recommended approaches are demonstrated herein.

Once an emerging FET has been systematically parameterized, benchmarking tables and plots are extremely useful for comparing devices from different reports. Because the electric fields are the driving forces within FETs, benchmarking performance metrics based on electric fields is natural. However, special care is needed in considering electric fields in devices, because they are spatially non-uniform and depend on many other factors, such as fringing fields and quantum capacitance $C_q$. Thus, the electric fields in nanoscale FETs are more complicated than the simple definition of an applied voltage divided by a physically defined length. For example, it is a reasonable assumption that the channel electric field in the channel, $E_{ch}$ increases linearly from source to drain in the linear region of operation, but $E_{ch}$ peaks sharply at the drain end of the channel in classical pinch-off (saturation)[30]. To account for this, the *average* $E_{ch}$ can be approximated as $(V_{DS} - 2I_{lin}R_c)/L_{ch}$ in the linear regime, accounting for voltage dropped at the contacts.

The vertical electric field at the source end, $E_{gate}$, can be estimated as $V_{OV}/t_{ins}$ if a planar gate is used. In turn, $E_{gate}$ and the gate insulator permittivity determine the carrier density in the channel. Yet, both $E_{gate}$ and the gate insulator permittivity are rather challenging to measure accurately. One more word of caution is justified: because many low-dimensional materials exhibit a low density of states, $C_{ins}$ needs to be replaced by $C_{ins}C_q/(C_{ins}+C_q)$ where quantum capacitance ($C_q$) can be approximated as $q^2$DOS (density of states)[31,32]. Only for $C_{ins} \ll C_q$ this expression becomes equivalent to $C_{ins}$. Because multiple parameters in Table 1 depend on $n_S$, benchmarking these versus $n_S$ is recommended to evaluate devices from different studies. A suggested list of benchmarking plots to evaluate device parameters and performance metrics is given in Table 2.

**Table 2| Suggested benchmarking plots for evaluating device performance compared with other FETs**

| Parameter | Benchmarking Plot | Notes |
|---|---|---|
| $I_{min}$ and $I_{max}$ | $I_{min}$ vs. $I_{max}$ | • Specify $L_{ch}$ or $L_{ch}$/EOT <br> • Ideally specify the carrier density at which $I_{max}$ is extracted (Note S1) |
| $I_{sat}$ | $I_{sat}$ vs. $n_S$ | Label $L_{ch}$ and $t_{ch}$ to imply channel resistance |
| $I_D$ | $I_D$ vs. $L_{ch}$ | At certain $V_{DS}$ and $n_S$ (e.g., $V_{DS}$ = 1 V and $n_S$ = $10^{13}$ cm$^{-2}$) |
| $R_c$ | $R_c$ vs. $n_S$ | • Or benchmark $R_c$ vs. $t_{ch}$ at certain $n_S$ <br> • Specify if semiconductor in contact regions is gated or not |
| $\mu_{con}$ | $\mu_{con}$ vs. $t_{ch}$ | For the same material, a thicker channel could have higher mobility due to less surface scattering |
| $g_m$ | $g_m$ vs. $n_S/L_{ch}$ | • Since $g_m = \frac{W}{L}\mu C_{ins}V_{ov}$, $g_m \propto C_{ins}V_{ov}/L_{ch}$ <br> • $n_S \approx C_{ins}V_{ov}/q$ near the source end, so $g_m \propto n_S/L_{ch}$ |
| SS | SS vs. $C_{ins}$ | • Larger $C_{ins}$ can yield smaller SS <br> • Identify Schottky barrier branch and thermal branch <br> • Or plot SS vs. $L_{ch}$ to show short-channel robustness <br> • Or plot SS vs $\log_{10}(I_D)$ |

Drain current is the key output of a FET and is also frequently benchmarked and compared. However, many comparisons are oversimplified and not fairly conducted as the drain current depends on many parameters. As mentioned in Table 2, we recommend benchmarking $I_D$ vs. $L_{ch}$ at certain $V_{DS}$ and $n_S$ values, enabling fair comparison between devices having different channel lengths. On the other hand, if a record $I_{sat}$ is claimed, we recommend benchmarking the maximum $I_{sat}$ vs. $n_S$ because it is a much closer indicator for the eventual drive current and ultimately sets the operating delay of a circuit stage (delay $\tau \propto CV_{DD}/I_{sat}$). As mentioned previously, assuming limited short-channel effects, $I_{sat}$ mainly depends on $n_S$ and not on $E_{ch}$. Usually, one performance metric depends on multiple parameters; hence, key parameters should be annotated on the benchmarking plot (see Table 2).

**Reporting and benchmarking example**

To demonstrate reporting and benchmarking based on the principles proposed above, MoS$_2$ is chosen as the example emerging channel material because it is among the most studied semiconducting nanomaterials in recent years and represents a family of 2D materials that holds promise for future transistor applications. Fig. 2a shows an example transistor based on monolayer (1L) MoS$_2$ grown by chemical vapor deposition (CVD). The device is top-contacted and back-gated, which is the most common and convenient FET structure used to explore emergent channel materials. The approach is as follows:

**Step 1**. The structural parameters of the device are determined and labelled (Fig. 2a). In this example, the gate insulator is $AlO_x$, which is grown by atomic layer deposition (ALD) with the oxide capacitance ($C_{ins} \approx 280$ nF/cm$^2$) evaluated from a capacitance-voltage measurement of a large-area test capacitor. The thickness of the oxide ($t_{ins} \approx 20$ nm) is further confirmed by cross-sectional transmission electron microscope (TEM) imaging. From the thickness and capacitance, the dielectric constant of the oxide is estimated to be $\varepsilon_{ins} \approx 6$. Other dimensions such as $L_{ch}$, $W_{ch}$, and $L_c$ are confirmed by scanning electron microscopy (SEM) after electrical characterization.

**Step 2**. $I_D$-$V_{GS}$ and $I_D$-$V_{DS}$ characterization are performed, making sure that $V_{DS}$ and $V_{GS}$ are swept high enough for the device to reach saturation, and the $V_{GS}$ sweep range is sufficient to observe $I_{min}$ in both the linear (low $V_{DS}$) and saturation (high $V_{DS}$) operation regions. $I_{max}/I_{min}$ of ~$4\times10^7$ at $V_{DS} = 4$ V can be extracted from the subthreshold curve in Fig. 2b. $I_{max}$ is extracted at $n_S \approx 1.4\times10^{13}$ cm$^{-2}$. $I_{min}$ is extracted under subthreshold conditions, where $V_{GS} < V_T$, yielding a negative $V_{OV}$ and $n_S \approx 0$. The larger hysteresis for $V_{DS} = 4$ V in Fig. 2b,c highlights the impact of the larger source-drain field on the interface charges in the channel. Due to hot-carrier stress from the high $V_{DS}$ (explained later), $V_T$ increases for high $V_{DS}$, resulting in an extracted DIBL of −274 to −436 mV/V considering the effect of hysteresis. Also, from the transfer curves in Fig. 2c, the maximum $g_m$(sat) and $g_m$(lin) are estimated to be ~ 59 µS/µm and ~1.7 µS/µm, respectively. In Fig. 2d, approximate current saturation is observed with $I_{sat}$ around 325 µA/µm obtained at $n_S \approx 1.3\times10^{13}$ cm$^{-2}$ at $V_{DS} = 4$ V. The blue and red shows linear and saturation regions, approximately. The $I_D$-$V_{DS}$ spacing is sub-linear at the highest $V_{GS}$, which is a sign of possible self-heating.

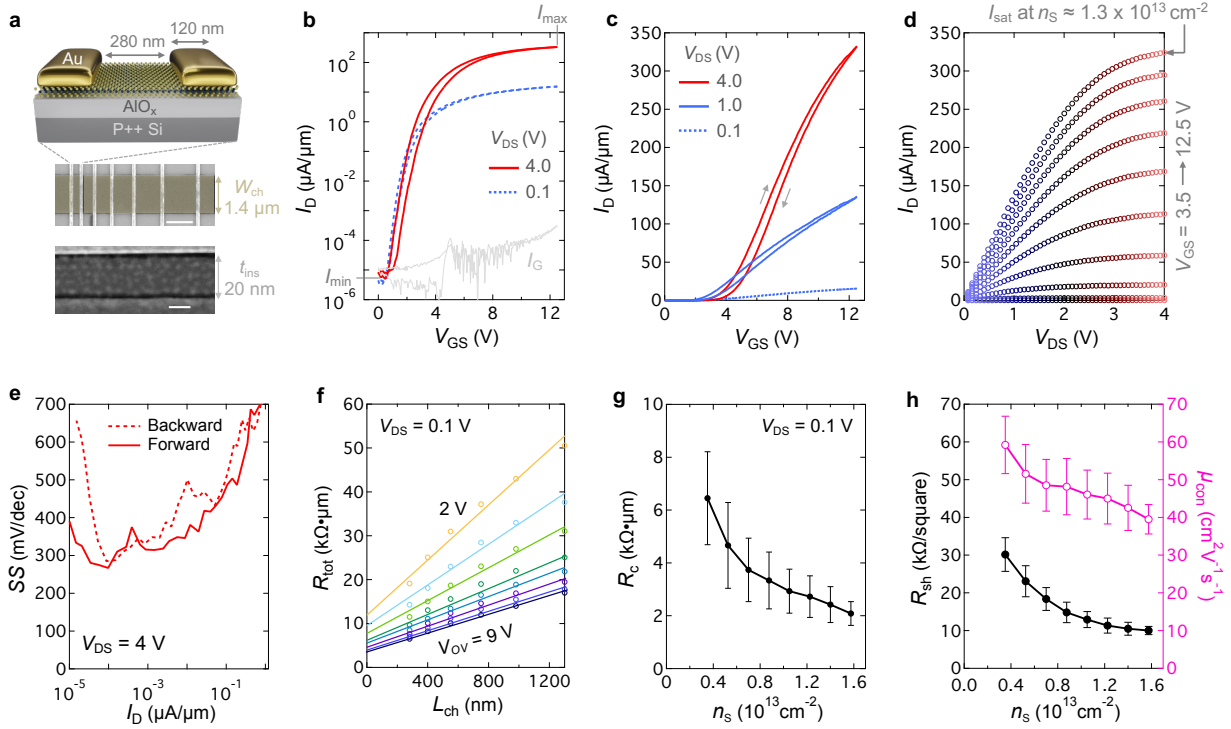

**Fig. 2 | Example of reporting device performance for monolayer Au-contacted MoS$_2$ FETs. a,** Device schematic and the basic structural parameters of the MoS$_2$ FET. An SEM image of a TLM structure with false-colored MoS$_2$ area (scale bar, 1 µm) and a cross-section TEM image of the AlO$_x$ (scale bar, 10 nm). **b,** Subthreshold ($I_D$-$V_{GS}$) curve of the device with $I_{max}$ and $I_{min}$ labelled. The $I_G$ is the gate leakage current at $V_{DS}$ = 4 V. The curve for $V_{DS}$ = 1 V is intentionally not shown so the plot is less crowded. **c,** Transfer ($I_D$-$V_{GS}$) curves of the device showing larger hysteresis with larger $V_{DS}$. **d,** Output ($I_D$-$V_{DS}$) curves of the device. $V_{GS}$ changes in steps of 1 V. **e,** $SS$ vs. $I_D$ for both forwards and backwards $V_{GS}$ sweeps in (**b**). **f,** Extraction of $R_c$ from the TLM structure in (**a**). The $R_{tot}$ is extracted at $V_{DS}$ = 100 mV. **g,** Contact resistance vs. $n_S$ showing the contact gating effect that is a result of the device operating as a Schottky barrier transistor with the gate modulating the semiconductor in the source/drain contact regions[33]. **h,** Extracted sheet resistance (left axis) and conductivity mobility (right axis) versus $n_S$. In **g** and **h**, error bars reflect 90% confidence interval from a least-squares fit of the TLM.

Figures 2a-d are used for primary characterization of one device, and more derived plots are shown in Fig. 2e-h, providing a more complete picture of the device characteristics. The device spread and parameter variations based on ten similar TLM structures are shown in Fig. S5. The full range of $SS$ vs. $I_D$ is plotted in Fig. 2d, with a minimum $SS$ of 280 mV/dec extracted in the subthreshold regime. Additionally, because $V_T$ depends on $V_{DS}$, it is key to extract $V_T$ at the associated $V_{DS}$ (as noted in Table 1). $R_c$, $R_{sh}$, and subsequently $\mu_{con}$ are extracted by using a TLM structure as shown in Fig. 2f-h. The $R_c$ is estimated to be around 2.1 kΩ·µm, which is comparable to the $R_{ch}$ of 2.8 kΩ·µm for the device with channel length of 280 nm. The relation between $n_S$ and extracted $R_c$ is plotted in Fig. 2g to show the effect of the overall back gate on the contact resistance (i.e., contact gating[33–35]). Fig. 2h shows

$\mu_{con}$ decreases from 59 to 40 cm$^2$V$^{-1}$s$^{-1}$ with increasing $n_S$, likely due to the increased electron scattering with the oxide surface roughness.

***Step 3***: As a simplified example, we benchmark key device performance parameters in Fig. 3 with a limited number of reports included. Currently, most papers do not report $I_D$-$V_{GS}$ at $V_{DS}$ (sat) or close to $V_{DD}$ as recommended above. Hence, plotting $I_{min}$ vs. $I_{max}$ at certain $V_{DS}$ (*e.g.*, $V_{DS}$ = 1 V) while annotating $L_{ch}$ is an acceptable approach (Fig. 3a). The upper limit of the carrier density is set at $n_S$ = $10^{13}$ cm$^{-2}$, ensuring a fair comparison of $I_{max}$. The $I_{max}/I_{min}$ ratio annotated on the right axis is also shown in the dashed lines in Fig. 3a. Due to a better electrostatic control from the gate, devices with a larger $L_{ch}$/EOT ratio tend to yield a higher $I_{max}/I_{min}$ (where EOT is the equivalent oxide thickness). Other parameters also play a role such as the leakage currents through the gate insulator or from source to drain. Large channel width can also produce a more accurate width-normalized $I_{min}$, especially when $I_{min}$ is below the instrument noise floor. For example, Illarionov *et al.*[36] demonstrated a relatively high $I_{max}/I_{min}$ ratio in devices with a 20 µm channel width. Further study is still needed to investigate how to achieve high $I_{max}$ and small $I_{min}$ in aggressively scaled devices (small $L_{ch}$, $L_c$, and EOT).

As mentioned previously, a high drain current in the saturation regime is a key performance metric. In Fig. 3b, $I_{sat}$ is plotted versus $n_S$ from representative studies of FETs based on MoS$_2$ as the channel material. We note that the $I_{sat}$ is extracted at different $V_{DS}$ because different devices have different channel lengths, as annotated in Fig. 3b. We caution against plotting $I_{sat}/L_{ch}$ vs. $n_S$ because it implies that $L_{ch}$ is the main limiting factor for $I_{sat}$, which is not necessarily true, especially for scaled devices where contact resistance typically dominates $I_{sat}$ performance. Importantly, it is clear that $I_{sat}$ needs to be further improved to meet the high-performance target of the most recent technology guidelines (at $V_{DD} \approx 0.65$ V near 2030)[37]. Many reports already used high $n_S$ but fell short on $I_{sat}$ even with channel lengths down to ~10 nm[38], being strongly limited by their contacts.

Recently, semimetal contacts such as bismuth have been shown to produce high-quality contacts[39]. It is nevertheless noteworthy that the two Bi-1L MoS$_2$ devices have a wide range of $I_{sat}$ performance, encompassing all the other devices in Fig. 3b, yet the channel length difference between the two devices is only 115 nm. Interestingly, one of the Bi-contacted devices ($L_{ch}$ = 150 nm) actually underperforms other Au-contacted devices with longer channel lengths. Hence, although some approaches show potential to achieve the International Roadmap for Devices and Systems (IRDS) high-performance goal of $I_{sat}$ for the post-2030 era[37], further investigation is still needed to reliably and reproducibly realize high $I_{sat}$ from a monolayer channel. In next-generation FETs, $I_{sat}$ could also be increased by shifting to nanosheet device designs, which stack multiple channels vertically to improve current density in the same device footprint[40].

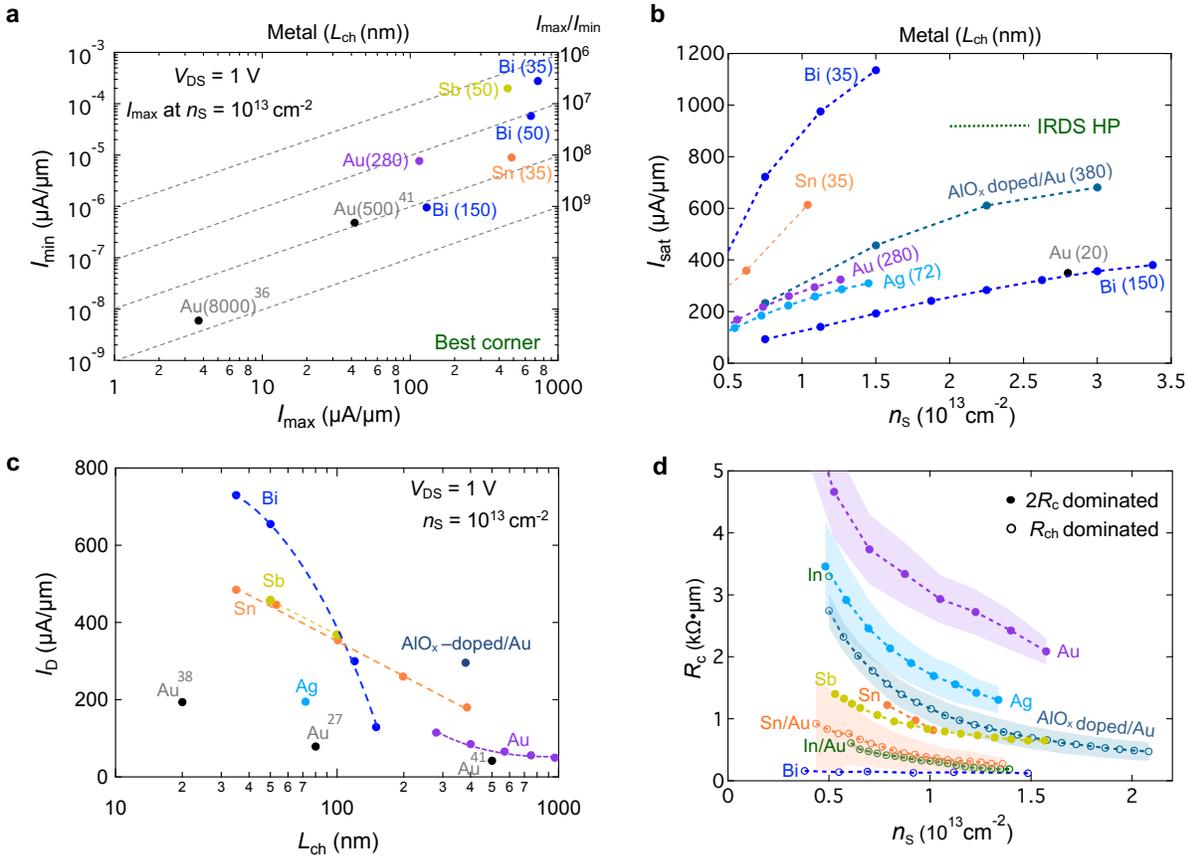

**Fig. 3 | Example Benchmarking device performance of monolayer MoS₂ FETs**. **a**, Benchmarking $I_{min}$ vs. $I_{max}$. The $I_{max}$ is extracted at $n_S = 10^{13}$ cm$^{-2}$. For simplicity, the $I_{min}$ of the data points are extracted at their respective smallest current. More rigorous benchmarking of $I_{min}$ vs $I_{max}$ is in Note S1. **b**, Benchmarking $I_{sat}$ versus $n_S$, where the channel length in nm is labelled next to the contact metals used in the devices. $I_{sat}$ are extracted at different $V_{DS}$ (listed in Table 3). These devices do not have the same $V_{DS}$(sat) as devices with different channel lengths and saturate at different $V_{DS}$. The IRDS HP is shown in a range of $n_S$ to represent uncertainties of the carrier density in future generation technologies. **c**, Benchmarking $I_D$ versus $L_{ch}$ at $V_{DS} = 1$ V and $n_S = 10^{13}$ cm$^{-2}$. Ref. [41] uses top-gate, while other reports use back-gate. **d**, Benchmarking $R_c$ versus $n_S$ in a few representative reports. The shaded regions represent uncertainties reported in the respective studies (Ag[14] uses 95% confidence interval, while In/Au[42], Sn/Au[42], and Au use standard error from the linear regression of TLM). The $R_c$ vs. $n_S$ of Sb[43] is obtained from Y-function method. The filled and open symbol show that the shortest channel device in the TLM structure is $2R_c$ and $R_{ch}$ dominated, respectively. If $R_{ch}$ dominates or the $R_c$ is over an order of magnitude smaller than the $R_{tot}$ for the smallest device in TLM, the extracted $R_c$ is of questionable validity; i.e., solid symbol data is more reliable per Note S3. Different colors are assigned to different reported devices. The purple Au data denote devices in Fig. 2. Most of the data are extracted from published reports: Ag[14], AlO$_x$ doped/Au[15], In/Au[42], Sn/Au[42], Sb[43], In[44] and Sn[45]. In **a** and **c**, reference numbers are added for Au-contacted devices to better differentiate their performance. A few studies are plotted as dotted lines to highlight the trends and to improve the clarity of the plots.

In some reports, a proper saturation current is not given. Also, since different devices use different channel lengths, $I_{sat}$ is often extracted at different $V_{DS}$. To highlight the impact of channel length, we recommend benchmarking $I_D$ vs. $L_{ch}$ at $V_{DS} = 1$ V and $n_S = 10^{13}$ cm$^{-2}$. This plot enables a direct comparison of devices with similar $L_{ch}$. In Fig. 3c, with channel length decreasing from 200 nm to 38 nm, both Bi[39] and Sn[45] contacted devices yield relatively large increases in drain current. We note that the Bi-contacted devices are based on different MoS$_2$ films. The different quality of the MoS$_2$ may partially contribute to the large increase of $I_D$ with a relatively small change in $L_{ch}$. Nevertheless, Figures 3c,d present the potential of atomic thin materials for producing high drain current, especially for scaled devices.

In Fig. 3d, $R_c$ is plotted against $n_S$ considering most devices have gated contacts (i.e., the back-gate modulates the channel and contacts[33–35]). While some reported $R_c$ values reach below 500 Ω·μm, their TLM extractions are all based on channel resistance-dominated devices, which can lead to questionable validity in their claimed $R_c$ (an artificially small or even a negative $R_c$ can be extracted, see Note S3 for details). We advocate that if a record $R_c$ is claimed from TLM, $R_{sh}$ should be cross-examined by using other methods such as four-probe measurements, which can provide a relatively accurate estimation of $R_{sh}$ vs. $n_S$ (Note S3). With $R_{sh}$ vs. $n_S$ from four-probe measurements, $R_c$ vs. $n_S$ can be derived by deducting the $R_{ch} = R_{sh}L_{ch}$ from $R_{tot}$ to confirm the TLM extracted value. Showing extraction of $R_c$ from many TLM structures can also increase confidence in the data by providing an average value of the $R_c$ rather than just the minimal value from a single TLM[46,47]. Furthermore, $R_c$ is heavily impacted by contact gating, as evident from the similar trends of $R_c$ vs. $n_S$ observed in different studies; therefore, further research is needed to obtain a small $R_c$ without gating the contacts.

Looking forward, many opportunities remain to develop transistors that simultaneously have small contact resistance, high $I_{sat}$, large $I_{max}/I_{min}$ ratio, and minimal short-channel effects by using emergent nanomaterials. To achieve this technological goal, interface engineering at the contacts and gate dielectric needs to be further investigated[48–50], along with progress in material synthesis[51] and integration[52]. Moreover, it is important to focus on channel thicknesses below ~3 nm, where low-dimensional nanomaterials can excel compared to Si, which suffers from poor carrier transport properties and a widened band gap in this thickness regime[53].

In addition to the example benchmarking plots in Fig. 3, other benchmarks can also be used to compare different devices. For example, as included in Table 2, plotting $I_{max}/I_{min}$ vs. $t_{ch}$ or $I_{max}/I_{min}$ vs. $L_{ch}$/EOT to compare the off-state device performance. Plotting $SS$ vs. $C_{ins}$ or $SS$ vs. $\log_{10}(I_D)$[54] can be used to evaluate subthreshold behaviour and trends across different devices. Finally, to show the quality of the channel materials, the channel sheet resistance or conductivity mobility can be plotted vs. carrier densities, as in Fig. 2h. Representative reports with relatively large $I_{sat}$ are listed in Table S1, including results for FETs with both monolayer and multilayer $MoS_2$ channels (example benchmarking plots in Note S6). In the literature, notable benchmarking examples can be found in Refs.[39,55] that highlight different channel materials, and in Ref. [56] that focuses on device performance in integrated circuits (speed, gain, density, power consumption, and fan-out capabilities, etc.).

**Conclusions**

Our guidelines should help put key performance metrics in a proper context and enable researchers to effectively report and benchmark emergent transistors based on various emergent nanomaterials. While each of the listed metrics is significant, it is not necessary — nor always possible — to extract and present all of them. As such, it is essential to completely describe the device geometry, to collect and report appropriate current-voltage characteristics, and to describe in detail the procedures followed in the experiments. The approaches used to analyze data and extract benchmarking metrics should also be described in detail. Depending on the context and need, we recommend three sets of parameters to report and benchmark. The first includes maximum saturation current, on/off-current ratio, transconductance, and subthreshold swing. These values can be directly obtained from the measured $I$-$V$ characteristics that cover both linear and saturation regimes. These values are mainly determined by the intrinsic material properties, gate stack configuration, and contact quality. The second includes the derived parameters such as mobilities and contact resistance, where uncertainty and statistical spread on these derived parameters should be shown. The third set of parameters are those specific to certain transistor demonstrations based on the target application[57,58]. For example, DIBL is essential when reporting and evaluating ultra-scaled FETs. It is important — whenever possible — to benchmark against other novel materials and also state-of-the-art in mature technology[39,55]. By using these guidelines, it should be possible to comprehensively and consistently reveal, highlight, discuss, compare, and evaluate device performance, thus helping to identify advances and opportunities in the search for improved transistors.

[1]Nanoscale Device Characterization Division, National Institute of Standards and Technology, Gaithersburg, MD 20899, USA. [2]Department of Electrical and Computer Engineering, Purdue University, West Lafayette, IN 47907, USA. [3]Department of Chemistry and Biochemistry, University of California, Los Angeles, Los Angeles, CA 90095, USA. [4]Theiss Research, La Jolla, CA 92037, USA. [5]School of Integrated Circuits, Peking University, Beijing 100871, China. [6]Electrical and Computer Engineering, New York University, Brooklyn, NY 11201, USA. [7]Center for Quantum Phenomena, Physics Department, New York University, New York, NY 10003, USA. [8]IMEC, Leuven, Belgium. [9]RWTH Aachen University, Chair of Electronic Devices, Otto-Blumenthal-Str. 2, Aachen 52074, Germany. [10]AMO GmbH, Advanced Microelectronic Center Aachen, Otto-Blumenthal-Str. 25, Aachen, 52074 Germany. [11]Key Laboratory for the Physics and Chemistry of Nanodevices and Center for Carbon-based Electronics, Department of Electronics, Peking University, Beijing, China. [12]Department of Electrical and Computer Engineering, University of Minnesota, Minneapolis, MN 55455 USA. [13]Department of Electrical Engineering, Stanford University, Stanford, CA 94305, USA. [14]Department of Electrical and Computer Engineering, Duke University, Durham, NC 27708, USA. [15]Department of Chemistry, Duke University, Durham, NC 27708, USA. †: corresponding authors.

**Acknowledgements**

We acknowledge Dr. Huairuo Zhang and Dr. Albert Davydov from the National Institute of Standards and Technology for their help on the TEM images of the oxide in Fig. 2. We acknowledge Dr. Guoqing Li and Prof. Linyou Cao from North Carolina State University for providing the CVD $MoS_2$ film. This work is supported by NEWLIMITS, a center in nCORE, a Semiconductor Research Corporation (SRC) program sponsored by NIST through award number 70NANB17H041. A.D.F acknowledges the support from the National Science Foundation under Grant ECCS 1915814. M.C.L. acknowledges funding from the European Union's Horizon 2020 research and innovation program under grant agreements 881603 (Graphene Flagship), 952792 (2D-EPL) and 829035 (QUEFORMAL), as well as the Deutsche Forschungsgemeinschaft (DFG, German Research Foundation) through the grants LE 2440/7-1 and LE 2440/8-1. Furthermore, support by the Bundesministerium für Bildung und Forschung (BMBF, German Ministry of Education and Research) through the grants 03XP0210 (GIMMIK) and 03ZU1106 (NeuroSys) is acknowledged. L.-M.P. acknowledges the National Science Foundation of China under the grant 61888102. S.J.K. acknowledges support from the NSF through award no. DMR-1921629. Fabrication and measurements were partially performed at the NIST Center for Nanoscale Science and Technology and at Duke Shared Manufacturing and Instrument Facility (SMIF).

**Author contributions**

All authors contributed to the preparation of the manuscript.

**Notes**

The authors declare no competing financial interest.

Certain commercial equipment, instruments, or materials are identified in this paper to specify the experimental procedure adequately. Such identifications are not intended to imply recommendation or endorsement by the National Institute of Standards and Technology (NIST), nor it is intended to imply that the materials or equipment identified are necessarily the best available for the purpose.

**Corresponding authors**

Zhihui Cheng, zhihui.cheng@alumni.duke.edu

Aaron D. Franklin, aaron.franklin@duke.edu

Curt A. Richter, curt.richter@nist.gov

**Data availability**

The data used in the paper are available from the corresponding authors upon reasonable request.

# Supplementary Information for
# How to Report and Benchmark Emerging Field-Effect Transistors

Zhihui Cheng[1,2,†], Chin-Sheng Pang[2], Peiqi Wang[3], Son T. Le[1,4], Yanqing Wu[5], Davood Shahrjerdi[6,7], Iuliana Radu[8], Max C. Lemme[9,10], Lian-Mao Peng[11], Xiangfeng Duan[3], Zhihong Chen[2], Joerg Appenzeller[2], Steven J. Koester[12], Eric Pop[13], Aaron D. Franklin[14,15,†], Curt A. Richter[1,†]

[1]Nanoscale Device Characterization Division, National Institute of Standards and Technology, Gaithersburg, MD 20899, USA. [2]Department of Electrical and Computer Engineering, Purdue University, West Lafayette, IN 47907, USA. [3]Department of Chemistry and Biochemistry, University of California, Los Angeles, Los Angeles, CA 90095, USA. [4]Theiss Research, La Jolla, CA 92037, USA. [5]School of Integrated Circuits, Peking University, Beijing 100871, China. [6]Electrical and Computer Engineering, New York University, Brooklyn, NY 11201, USA. [7]Center for Quantum Phenomena, Physics Department, New York University, New York, NY 10003, USA. [8]IMEC, Leuven, Belgium. [9]RWTH Aachen University, Chair of Electronic Devices, Otto-Blumenthal-Str. 2, Aachen 52074, Germany. [10]AMO GmbH, Advanced Microelectronic Center Aachen, Otto-Blumenthal-Str. 25, Aachen, 52074 Germany. [11]Key Laboratory for the Physics and Chemistry of Nanodevices and Center for Carbon-based Electronics, Department of Electronics, Peking University, Beijing, China. [12]Department of Electrical and Computer Engineering, University of Minnesota, Minneapolis, MN 55455 USA. [13]Department of Electrical Engineering, Stanford University, Stanford, CA 94305, USA. [14]Department of Electrical and Computer Engineering, Duke University, Durham, NC 27708, USA. [15]Department of Chemistry, Duke University, Durham, NC 27708, USA. †: corresponding authors.

Note S1: Rigorously reporting and benchmarking $I_{max}/I_{min}$

Note S2: Different methods to extract threshold voltage $V_T$ and its uncertainties

Note S3: Extracting $R_c$ from TLM

Note S4: Additional Parameters

Note S5: Demonstration of device spread and parameter variations

Note S6: Benchmarking devices with different channel thicknesses

**Note S1: Rigorously reporting and benchmarking $I_{max}/I_{min}$**

The $I_{max}/I_{min}$ ratio is the most-used figure of merit for evaluating off-state performance of a transistor and it depends on multiple factors. First, $I_{max}/I_{min}$ is impacted by material properties such as material quality, bandgaps, and doping conditions. Secondly, the specific device structure and geometry, including the channel length and gate capacitance, also affect the achievable $I_{max}/I_{min}$. Thirdly, the same device operating at different operation regimes might also have different $I_{max}/I_{min}$. Combined with the absence of a well-defined $V_{DD}$, these factors make rigorous reporting and benchmarking guidelines strongly desired for $I_{max}/I_{min}$. Rigorously reporting $I_{max}/I_{min}$ involves clearly identifying the $V_{DS}$ and $V_{GS}$ values at which the $I_{max}$ and $I_{min}$ are extracted. This practice will help fairly benchmark $I_{max}/I_{min}$ between different devices.

Similar to Fig. 3a, Fig. S1a represents benchmarking $I_{max}/I_{min}$ at $V_{DS} = 1$ V, which is a commonly used bias condition when obtaining subthreshold curves. Under $V_{DS} = 1$ V, short-channel devices (e.g., Device A) may operate at the saturation regime, whereas long-channel devices (e.g., Device B) at the linear regime. While using the same $V_{DS} = 1$ V is a convenient and simple method, the $I_{max}/I_{min}$ likely represent different operation regimes for the two example devices. For a more precise comparison, instead of extracting the $I_{max}$ and $I_{min}$ at the same $V_{DS}$ value for the different devices, it is possible to ensure they are extracted in the same regime (linear or saturation) by using the drain-source electric field instead of $V_{DS}$; e.g., $E_{DS} = 0.1$ V/μm for the linear regime. This improved comparison is demonstrated in Fig. S1b, where the carrier density at which the $I_{max}$ is extracted is also labeled.

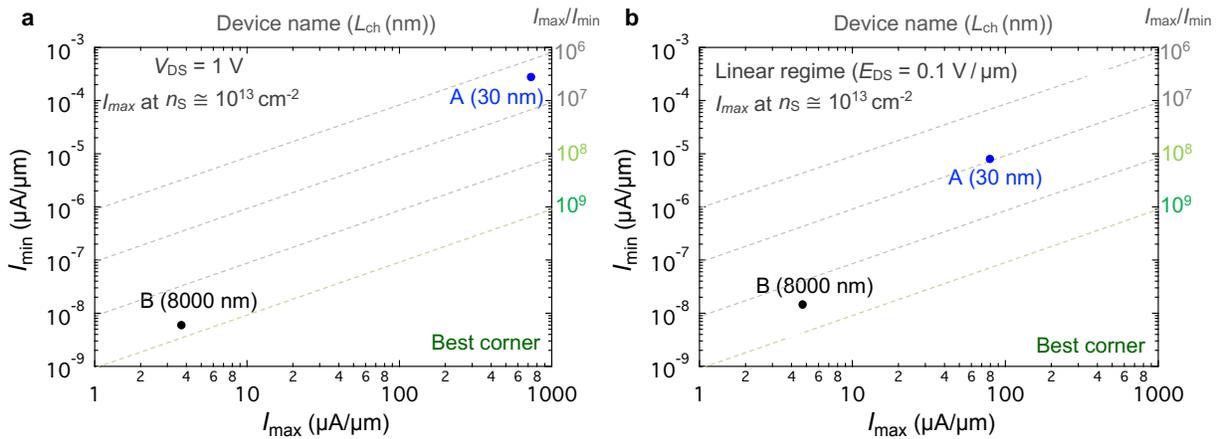

**Fig. S1 | Rigorously benchmarking $I_{max}/I_{min}$. a**, Benchmarking $I_{max}/I_{min}$ at $V_{DS} = 1$ V, similar to Fig. 3a. Devices A and B are not actual experimental data and are used to demonstrate benchmarking devices with different channel lengths. The carrier densities at which the $I_{max}$ and $I_{min}$ are extracted are also labeled. **b**, Benchmarking $I_{max}/I_{min}$ at the same operation regime, which is the linear regime with drain-to-source field $E_{DS} = 0.1$ V/μm used as an example. Note, the diagonal dashed lines indicate certain values of $I_{max}/I_{min}$.

Additional factors can impact $I_{max}/I_{min}$ ratio. When benchmarking devices with the same channel material but with different channel thicknesses, plotting $I_{max}/I_{min}$ vs. $t_{ch}$ is suggested. It is noteworthy that $I_{max}/I_{min}$ also depends on channel length and the equivalent oxide thickness (EOT) of the gate insulator. When the devices to be benchmarked have the same channel thickness, then benchmarking $I_{max}/I_{min}$ vs. $L_{ch}$/EOT should be considered. In addition, $I_{min}$ can be limited by the measurement instrumentation, especially in materials with larger band gap such as monolayer TMDs. Additional factors impacting $I_{max}/I_{min}$ include contact resistance (which limits $I_{max}$), leakage currents (which dominate $I_{min}$) and parasitic capacitances (which affect electrostatic control of the gate), etc.

**Note S2: Different methods to extract threshold voltage $V_T$ and its uncertainties**

*Different methods to extract $V_T$*

A standard method to extract the $V_T$ is not yet available. Refs. [4–7] have investigated and compared different $V_T$ extraction methods. Specifically, in Ref. [4], the authors investigated eleven approaches to extract the $V_T$ and found that similar $V_T$ values in the linear regime can be extracted. In Refs. [5,6], the authors proposed the Y-function method to eliminate the $R_c$ effect when evaluating mobility values. In their studies, the $V_T$ is extracted from the $I_D/g_m^{0.5}$ curve.

*Uncertainty of $V_T$ induced by I-V sweeps*

The threshold voltage $V_T$ uncertainties can propagate to the estimation of carrier density. Fig. S2 shows an example of $V_T$ shift due to different sweeps ($I_D$-$V_{GS}$ vs. $I_D$-$V_{DS}$) on the same device illustrated in Fig. 2c.

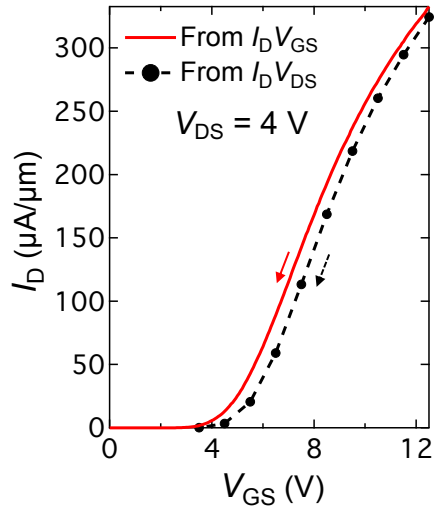

**Fig. S2** | Comparison of transfer curves at $V_{DS}$ = 4 V from $I_D$-$V_{GS}$ sweep (red) and extraction from $I_D$-$V_{DS}$ sweeps (black) based on data in Figures 2b,d of the main text. The direction of the sweeps is the same (backward). A $V_T$ shift of ~0.6 V is observed comparing the $I_D$-$V_{GS}$ and $I_D$-$V_{DS}$ sweeps. We presume this shift arises from hot-carrier stress during the $I_D$-$V_{DS}$ sweeping at high $V_{DS}$ and $V_{GS}$. Due to high electric fields, energetic electrons become trapped in the gate oxide as fixed charges[8], increasing the $V_T$. This phenomenon highlights the dependence of $V_T$ on $I_D$-$V_{DS}$ sweeping.

**Note S3: Extracting $R_c$ from TLM**

Extracting $R_c$ from TLM data can lead to significant uncertainties. As a reminder, TLM data is a plot of the total resistance ($R_{tot}$) from a set of devices having all things consistent with their structure and materials except their channel length. TLM data is plotted as $R_{tot}$ vs $L_{ch}$, where the $R_{tot} = R_{ch} + 2R_c = R_{sh}L_{ch} + 2R_c$, which means when the channel length is zero (y-axis intercept) $R_{tot} = 2R_c$. A reliable $R_c$ must be derived from TLM data that includes devices in the short-channel limit, where **$R_c$ dominates the total resistance** or at least is comparable to the $R_{ch}$. Ideally, the sheet resistance and its variation need to be sufficiently small; otherwise, a small variation in $R_{sh}$ could lead to substantial errors in $R_c$. Three examples of TLM data are presented in Fig. S3a, all claiming an $R_c$ of 500 Ω·µm, based on the intercepts to the $R_{tot}$ axis. However, due to the larger $R_{sh}$, the uncertainty of $R_c$ in TLM A is significantly larger than TLM B (e.g., a small change in $R_{sh}$ for TLM A would result in a large change in the extracted $R_c$). In TLM A, the shortest channel device has an $R_{tot}$ of 10 kΩ·µm, which is much larger than the claimed $R_c$ of 500 Ω·µm, whereas in TLM B, the $R_{tot}$ is 2 kΩ·µm and the claimed $R_c$ is 500 Ω·µm. As noted above, with a slight variation in the $R_{sh}$ in TLM A, the $R_c$ might be extracted as a very small (and often untrue) value or even a negative value.

Hence, it is critical to have both of the following in a TLM: 1) devices with channel length small enough to ensure $2R_c$ dominates $R_{tot}$ and 2) a sufficient number of longer channel length devices to ensure a reliable $R_{sh}$ extraction based on a linear fit. Another possible scenario to avoid is illustrated with TLM C, which represents a TLM data set where all of the devices are $2R_c$ dominated thus leading to an inaccurate estimation of $R_{sh}$. This comparison further highlights the need for proper TLM data; as a summary, a valid TLM data set should have at least four channels and include at least one each of contact and channel resistance-dominated devices.

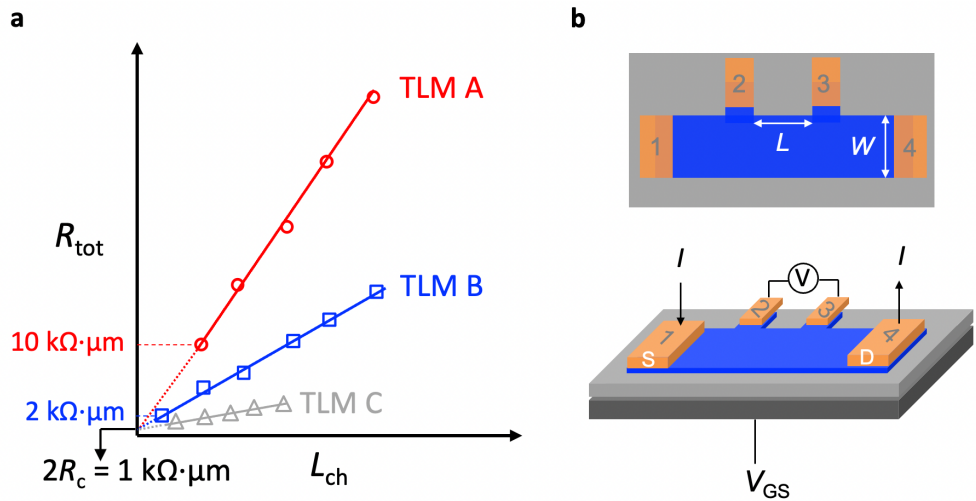

**Fig. S3** | **a**, Demonstration of TLM data sets with all $R_{ch}$-dominant devices (TLM A, leading to unreliable $R_c$ extraction), all $R_c$-dominant devices (TLM C, leading to unreliable $R_{sh}$ extraction), and an appropriate balance of devices (TLM B). **b**, Example diagram of four-probe measurements to extract $R_{sh}$, with a back-gate as a demonstration. The channel resistance $R_{ch}$ between probe 2 and 3 is $R_{14,23} = V_{23}/I_{14}$, where $I_{14}$ is the current flowing from probe 1 to 4 and $V_{23}$ is the voltage between probe 2 and 3. $R_{sh} = R_{14,23}(W/L)$. $R_{sh}$ can be obtained while $V_{GS}$ is swept, yielding $R_{sh}$ vs. $V_{GS}$ or carrier density, which can be used to cross-examine the $R_{sh}$ extracted from TLM.

Since there can be considerable variation in the quality of emerging semiconducting materials, even on the same chip (e.g., variation in crystal quality in $MoS_2$), this can translate to a sizeable variation in $R_{sh}$ from one device to another. Note, a perfect TLM is based on the assumption that $R_{sh}$ is the same for all devices being tested, so any disruption to this assumption translates to inaccuracy in the extraction of $R_c$. Hence, it is highly recommended to fabricate several TLM device sets and plot the full distribution of resultant TLM data (multiple data points for each channel length). This provides evidence for how dependable the extraction is; a good example of this is in Ref. [9] and Fig. S5g below.

Another critical aspect of proper TLM data sets is the need for reporting the uncertainties of $R_{sh}$ and $R_c$ in standard error or confidence interval from a least-squares fit of the TLM curve; an example of this can be seen in Fig.2h and Ref. [10]. Lastly, if a record parameter is claimed based on TLM, four-probe measurements[7,11] should ideally be used as a complementary method to cross-check the results. Traditionally, four-probe measurements are used to characterize the resistivity of "bulk" doped films and a gate bias is not involved. However, to estimate the $R_{sh}$ for emergent semiconducting materials, it is necessary to sweep the gate bias so that $R_{sh}$ vs. gate voltage (carrier density, $n_S$) can be obtained. An example of the four-probe measurement is given in Fig. S3b. It is key to have probe 2 and 3 placed to the side of the channel so that the voltage probes (2 and 3) do not obstruct the current flow from probe 1 to 4. After obtaining $R_{sh}$ vs. $n_S$, $R_c$ vs. $n_S$ can be derived by subtracting $R_{sh}$ from $R_{tot}$.

A final caution regarding the use of TLM data is regarding the transfer length ($L_T$), which is the length of the contact over which the majority of carrier injection occurs between the metal and semiconductor. Traditionally, $L_T$ was also extracted from TLM data as the intercept of the linear fit with the $L_{ch}$ (x-axis). However, extracting $L_T$ in this manner assumes that the sheet resistance of the semiconductor in the metal-contacted region is the same as for the semiconductor in the channel region. For emergent semiconducting materials, particularly low-dimensional materials like CNTs or 2D TMDs, this is not a valid assumption as transport through the materials happens predominantly (if not entirely) on the surface and is strongly dependent on the materials interfacing with the semiconductor. Because $L_T$ has significant implications for the scalability of the contact length in emergent FETs, it must not be improperly extracted and reported from simple TLM data sets. This is commented on further in the next section (Note S4).

**Note S4: Additional Parameters**

In the main text, we covered the most representative parameters for reporting and benchmarking emergent FETs. Here we describe some additional parameters used in specific studies.

*Normalization of $I_D$ for 1D devices*: For 1D or close to 1D devices, to compare $I_D$, it is sometimes necessary to convert the $I_D$ per CNT/nanowire/nanosheet stack to $I_D$ per µm. If the 1D channel materials are aligned CNTs or nanowires (Fig. S4a), the normalization of currents depends on the density of the 1D channel or the number of 1D channels in 1 µm. If the CNTs or nanowire are a dense network, treating the channel similar to a 2D material is more appropriate. For gate-all-around nanosheet devices (Fig. S4b), it is common to report $I_D$ per nanosheet stack. If reporting the $I_D$ per channel footprint, then the current per nanosheet stack should be divided by the nanosheet width ($W_{NS}$). A more detailed reporting of $I_D$ per channel width would require the calculation total channel width of the nanosheets in the stack, which is close to $W_{NS} \times T_{NS} \times$ number of nanosheets in the stack. The normalized drain currents per channel width is $I_D$ per stack divided by the total channel width of the stack. Ultimately, what matters is both an indication of what is achieved on a per nanomaterial/structure basis (e.g., per CNT or per nanosheet stack) as well as what is achieved on a per aerial footprint width basis (i.e., µA per $W_{ch}$).

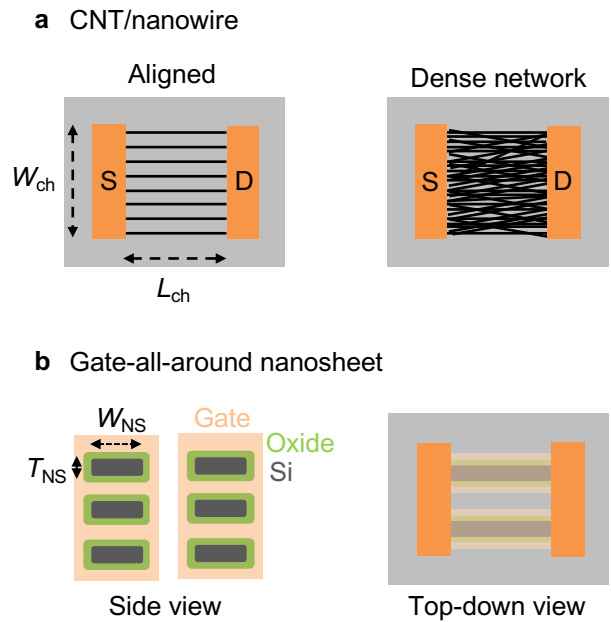

**Fig. S4** | Normalization of currents for transistors based on 1D or close to 1D channel materials. **a**, top-down view of transistors based on CNTs or nanowires in the form of aligned channels and a dense network. **b**, side-view and top-down (aerial) view of gate-all-around nanosheet transistors.

***Schottky barrier height***: For Schottky contacts between metal and semiconductors, the Schottky barrier height determines the efficiency of carrier transport in the contacts. Hence, properly extracting the barrier height is key for comparing different contact engineering approaches. Examples of Schottky barrier extraction can be found in Ref. [3,12–14], with some variances in the equations and approaches used.

***Transfer length*** ($L_T$): Because charge carriers tend to crowd near the contact edge when transported between the metal contacts and the channel material, only a certain portion of the contacts actively participate in the carrier transport process. The transfer length denotes the distance over which most of the current transfers in the contact. Traditionally, $L_T$ can be estimated as $\sqrt{\frac{\rho_c}{\rho_{sh}}}$ from the transfer length method plot, where the $\rho_c$ is the specific contact resistivity (unit: $\Omega \cdot cm^2$) and $\rho_{sh}$ is the sheet resistance underneath the contact (unit: $\Omega$/square) – this is discussed in some detail in Note S3 above. However, this estimation is not reliable for emergent transistors due to: 1) many of the emergent transistors use ultra-thin, low-dimensional nanomaterials, which can alter the current crowding behavior; 2) lots of research-grade emergent transistors have gated contacts, which complicates the estimation of the sheet resistance underneath the contact; and 3) the difference in the interface between the metal-semiconductor and gate insulator semiconductor leads to further differences in sheet resistance of the semiconductor in these regions. Hence, accurately determining $L_T$ will depend on physically scaled contacts to observe the contact scaling behavior.

***Interface trap density*** ($D_{it}$): This parameter is essential for studying and evaluating novel gate dielectrics. Different methods for determining $D_{it}$ can be found in Ref. [12].

***High-frequency response of the gate insulator capacitance***: Although most of the parameters covered in the main text are low-frequency parameters, for devices to be eventually used in high-frequency and high-performance applications, it is necessary to properly evaluate the high-frequency response of the gate insulator capacitance and the current-voltage characteristics[15].

**Note S5: Demonstration of device spread and parameter variations**

As indicated in the main text, we recommend showing device spread and parameter variations to demonstrate the full picture of the reported devices. In Fig. S5, based on ten TLM structures similar to the one used in Fig. 2, we demonstrate the spread of device characteristics and parameter variations using a cumulative distribution function plot and boxplots.

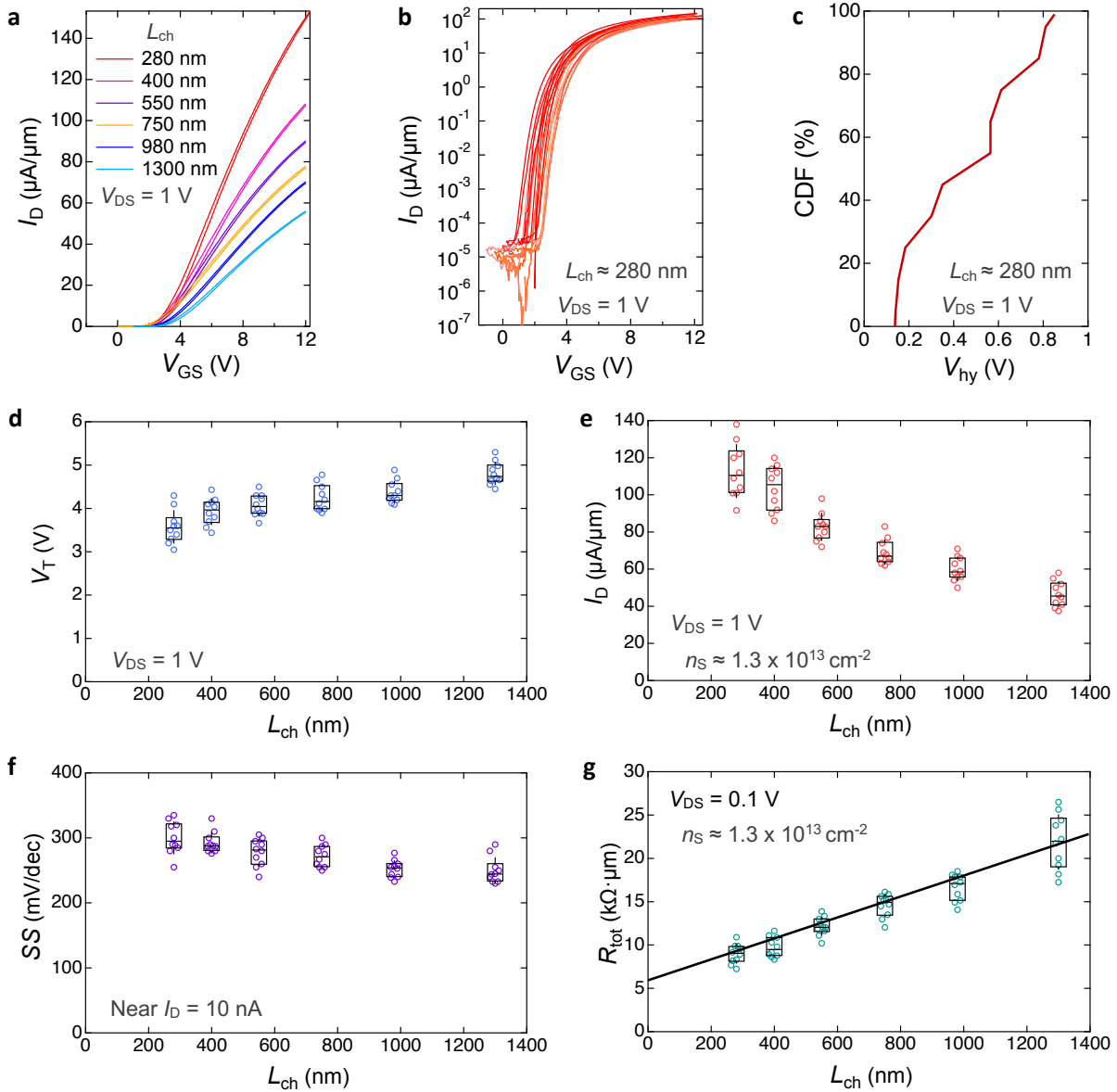

**Fig. S5** | Demonstration of device spread and parameter variations based on ten TLM structures. **a**, Example transfer curves ($I_D$-$V_{GS}$) for the TLM structure in Fig. 2. **b**, Measured subthreshold curves ($I_D$-$V_{GS}$) of ten devices with $L_{ch}$ of 280 nm of 280 nm. **c**, Cumulative distribution function (CDF) of the hysteresis voltage for the ten devices with $L_{ch}$ of 280 nm. Boxplots of (**d**) $V_T$, (**e**) $I_D$ at $V_{DS}$ = 1 V and $n_S$ = 1.3×10$^{13}$ cm$^{-2}$, (**f**) the minimal $SS$ extracted near $I_D$ = 10 nA, and (**g**) ten TLM plots with the median $R_c$ value of 3 kΩ·μm.

**Note S6: Benchmarking devices with different channel thicknesses**

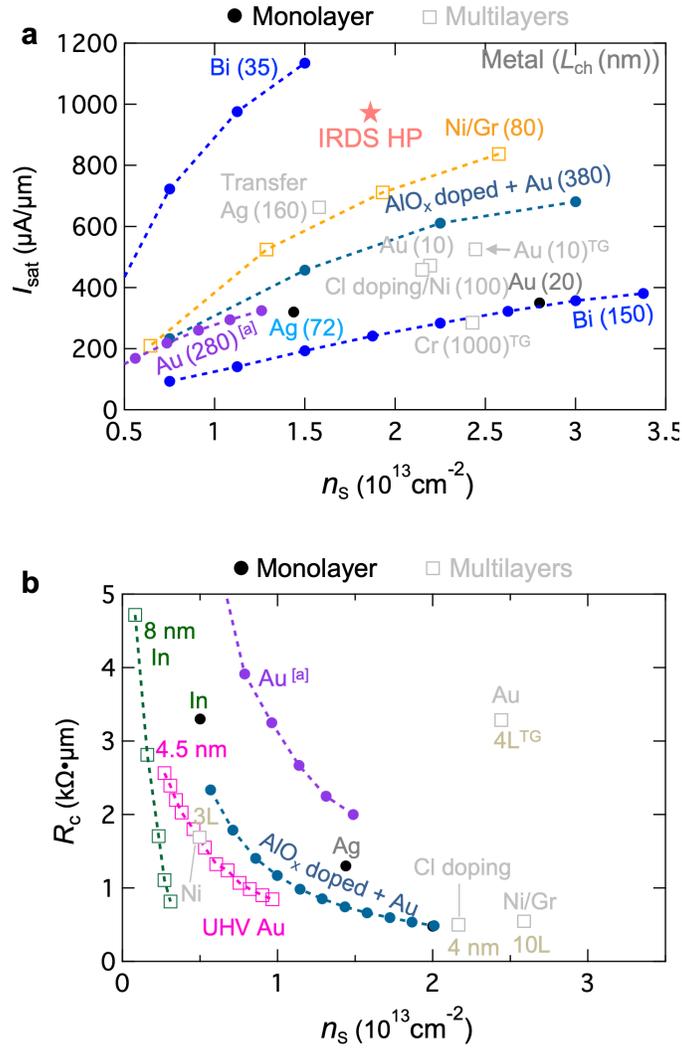

**Fig. S6** | Example benchmarking device performance of MoS$_2$ FETs with different channel thicknesses. **a**, Benchmarking $I_{sat}$ versus $n_S$, where the channel length in nm is labeled next to the metal contacts used in the devices. **b**, Benchmarking $R_c$ versus $n_S$ in a few representative reports, where the channel thickness is labeled for the multilayer channels. "a" stands for the example device in Fig. 2. A few studies are plotted as dotted lines to keep the plots from cluttering and also highlight the trends. Different colors are assigned to different reported devices. Most of the data are extracted from the following published reports: transferred Ag[16], AlO$_x$ doped+Au[17], Ag[18], In[19], Ni[20], Cl doping[1], Ni/Gr[2], and Bi[3].

Representative reports with relatively large $I_{sat}$ are listed in Table S1, with both monolayer and multilayer $MoS_2$ channels included.

**Table S1| Representative reporting on studies of $MoS_2$ FETs with $I_{sat}$**

| Contacts | Ref. | $t_{ch}$ (nm) | $L_{ch}$ (nm) | $n_S$ ($10^{13}$ cm$^{-2}$) | $R_c$ (kΩ·μm) | $V_{DS}$ (V) | $I_{sat}$ (μA/μm) |
|---|---|---|---|---|---|---|---|
| Ag | 18 | 1L | 72 | 1.44 | 1.3 | 1.8 | 320 |
| Au | a | 1L | 280 | 1.3 | 2 | 4 | 325 |
| Au | 21 | 1L | 20 | 2.8 | N/A | 2 | 350 |
| Bi | 3 | 1L | 150 | 3.4 | 0.12 | 2 | 380 |
| Ni+Cl doping | 1 | 4 | 100 | 2.16 | 0.5 | 1.6 | 460 |
| Au | 21 | 6 | 10 | 2.19 | N/A | 2 | 470 |
| Sn | 22 | 1L | 35 | 1 | 0.84 | 1.5 | 615 |
| Transferred Ag | 16 | 4~20 | 160 | 1.58 | N/A | 3 | 660 |
| $AlO_x$+Au | 17 | 1L | 380 | 2 | 0.48 | 5 | 700 |
| Ni/Gr | 2 | 10L | 80 | 2.58 | 0.54 | 2 | 830 |
| Ni | 23TG | 4.2 | 1000 | 2.43 | N/A | 3 | 290 |
| Au | 24TG | 1L | 10 | ~7 | 1.7 | 2 | 425 |
| Cr | 25TG | ~4L | 400 | 2.44 | 3.3 | 4 | 526 |

a: the example $MoS_2$ FET in Fig. 2a; TG = top gate; 1L = monolayer; 2L = bilayer, etc.
The Table lists $I_{sat}$ in ascending order but not including the last three rows because they are top-gated. All other FETs were back-gated. $\mu_{FE}$ is not benchmarked as several of the studies listed have overestimated values. More studies that may not have $I_{sat}$ reported can be accessed at Ref. [26].